\documentclass[a4paper]{article}

\usepackage{icrc2013}
\usepackage[english]{babel}
\usepackage{float}

\title{Recurring $^{3}$He-rich Solar Energetic Particle Events}

\shorttitle{$^{3}$He-rich SEP Events}

\authors{
R. Bu\v{c}\'ik$^{1}$,
D. E. Innes$^{1}$,
U. Mall$^{1}$,
A. Korth$^{1}$,
G. M. Mason$^{2}$
}

\afiliations{
$^1$ Max-Planck-Institut f\"{u}r Sonnensystemforschung, Katlenburg-Lindau, Germany \\
$^2$ Applied Physics Laboratory, Johns Hopkins University, Laurel, MD, USA
}

\email{bucik@mps.mpg.de}

\abstract{Using the SIT instrument aboard STEREO we have examined the abundance
of the $^{3}$He during the ascending phase of solar cycle 24 from January 2010 through
December 2012. We report on several cases when $^{3}$He-rich solar energetic particle
events were successively observed on ACE and STEREO-A with delays consistent with
the Carrington rotation rate. In the investigated period ACE and STEREO-A were
significantly separated in the heliolongitude corresponding to solar rotation
times of 5 to 10 days. We inspect STEREO-A EUV images and use the potential-field
source-surface extrapolations together with in-situ magnetic field data to identify
responsible solar sources. We find the $^{3}$He/$^{4}$He ratio highly variable in these events
and correlated between the spacecraft for the cases with the same connection region
on the Sun.}

\keywords{solar energetic particles, flares, abundances, PFSS.}

\begin{document}
\maketitle

\section{Introduction}

$^{3}$He-rich solar energetic particle (SEP) events are characterized by huge enhancements
of the rare isotope $^{3}$He over solar system abundances. The enrichment of
the $^{3}$He is believed to be caused by selective heating mechanisms
due to its unique charge to mass ratio (see e.g., review by \cite{bib:mason07}).
Recently the sources of such events have been investigated with help of
imaging observations \cite{bib:nitta,bib:wang}. The sources were small flares
located near open field region and showed tendency to recur \cite{bib:wang}.

There were reported a few simultaneous $^{3}$He-rich SEP events \cite{bib:reames,bib:wieden}
observed on spatially separated spacecraft (s/c). Multi-spacecraft observations
of successive events have not
been systematically investigated. In the present paper, we examine $^{3}$He-rich SEP events
successively observed on two s/c widely separated in heliolongitude and identify
cases where they are connected to the same solar active region. We use
observations from the ACE and STEREO-A (STA) s/c
which had a separation between 65$^\circ$ and 130$^\circ$ during the study period
(January 2010-December 2012).

\section{Observations}

Energetic $^{3}$He observations used in this paper are from time-of-flight
mass spectrometers SIT \cite{bib:mason08} on STA and ULEIS \cite{bib:mason98} on ACE.
The STA s/c is in a heliocentric orbit at $\sim$1 AU near the ecliptic plane
increasing its separation from Earth at a rate of $\sim$22$^\circ$/year.
The ACE s/c is in an orbit around the L1 point. Since SIT is less sensitive
to small $^{3}$He-rich SEP events than ULEIS we
identified events on STA first and then we search for the corresponding
events on ACE.

\begin{table*}[!htbp]
\begin{center}
\begin{tabular}{ c c c c c c c c c c}

\hline Year & \multicolumn{2}{c}{Start time} & \multicolumn{2}{c}{$^{3}$He/$^{4}$He$^\textsuperscript{a}$} & \multicolumn{2}{c}{Fe/O$^\textsuperscript{a}$} & \multicolumn{2}{c}{$^{3}$He Fluence$^\textsuperscript{a}$ (x$10^3$)} & AR\\
\cline{2-3} \cline{4-5} \cline{6-7} \cline{8-9}
       & ACE & STA & ACE & STA & ACE & STA & ACE & STA &\\ \hline
2010   & Jan 27.3   & Feb  2.8   & 0.13$\pm$0.02 & 0.29$\pm$0.04 & 0.58$\pm$0.10 & 0.57$\pm$0.21 & 5.72$\pm$0.83 & 7.78$\pm$0.91 & 1041\\
2010   & Feb  8.7$^\textsuperscript{b}$   & Feb 14.3   & 0.21$\pm$0.02 & 0.54$\pm$0.05 & 0.89$\pm$0.16 & 1.50$\pm$0.34 & 21.5$\pm$1.6  & 16.0$\pm$1.3  & 1045\\
2010   & Oct 17.9   & Oct 21.4   & 0.46$\pm$0.04 & 1.12$\pm$0.24 & 1.27$\pm$0.23 & 2.25$\pm$1.35 & 21.0$\pm$1.6  & 4.48$\pm$0.69 & 1112\\
2010   & Oct 17.9   & Oct 24.6   & 0.46$\pm$0.04 & 0.16$\pm$0.03 & 1.27$\pm$0.23 & 1.17$\pm$0.65 & 21.0$\pm$1.6  & 3.52$\pm$0.61 & 1112\\
2010   & Nov 14.6   & Nov 22.4   & 0.20$\pm$0.02 & 1.24$\pm$0.16 & 1.20$\pm$0.18 & 1.00$\pm$0.43 & 26.3$\pm$1.8  & 13.8$\pm$1.2  & 1124\\
2010   & Nov 17.7   & Nov 22.4   & 2.75$\pm$0.33 & 1.24$\pm$0.16 & 1.25$\pm$0.34 & 1.00$\pm$0.43 & 32.8$\pm$2.0  & 13.8$\pm$1.2  & 1124\\
2011   & Jan 27.9   & Feb  1.8   & 0.08$\pm$0.01 & 0.09$\pm$0.02 & 1.15$\pm$0.18 & 4.67$\pm$2.97 & 10.7$\pm$1.1  & 1.71$\pm$0.43 & 1149\\
2011   & Apr 29.8   & May  6.4   & 0.02$\pm$0.01 & 3.90$\pm$0.62 & ...           & 1.43$\pm$0.70 & 0.95$\pm$0.34 & 18.5$\pm$1.4  & 1197\\
2011   & Jul  9.0   & Jul 16.4   & 2.35$\pm$0.13 & 3.28$\pm$0.20 & 1.31$\pm$0.20 & 1.13$\pm$0.28 & 127$\pm$4     & 105$\pm$3     & 1246\\
2012   & Jan  7.3   & Jan 16.0   & 1.33$\pm$0.26 & 0.82$\pm$0.22 & ...           & 1.67$\pm$1.22 & 7.15$\pm$0.92 & 2.45$\pm$0.51 & 1392\\
2012   & Apr 24.9   & May  2.0   & 0.11$\pm$0.01 & 0.05$\pm$0.01 & 0.99$\pm$0.11 & 0.97$\pm$0.18 & 23.4$\pm$1.7  & 7.36$\pm$0.89 & 1461\\
2012   & Nov 20.5   & Nov 30.9   & 6.72$\pm$0.52 & 0.80$\pm$0.27 & 0.89$\pm$0.16 & ...           & 154$\pm$4     & 1.60$\pm$0.41 & 1613\\  \hline
\end{tabular}
\begin{tabular}{l}
\hspace{-85mm}
$^\textsuperscript{a}$320--450 keV/n; fluence units - particles (cm$^2$ sr MeV/n)$^{-1}$ \\
\hspace{-85mm}
$^\textsuperscript{b}$event discussed in \cite{bib:wieden} \\
\end{tabular}
\caption{Recurring $^{3}$He-rich SEP events consecutively detected by ULEIS/ACE and SIT/STEREO-A.}
\label{table_single}
\end{center}
\end{table*}

In the survey period there were 32 STA $^{3}$He-rich events found in the energy
range $0.2-0.5$ MeV/n (8 events in 2010; 12 in 2011 and 12 in 2012). In 19 cases
a $^{3}$He-rich event was observed at the same energy earlier by ACE with the start
time within $\pm$2 days
of corotation delay between the two s/c. A common connection solar region
was found in 12 cases, which are listed in Table \ref{table_single}.

 \begin{figure*}[!htbp]
 \vspace{-5mm}
  \centering
  \includegraphics[width=\textwidth]{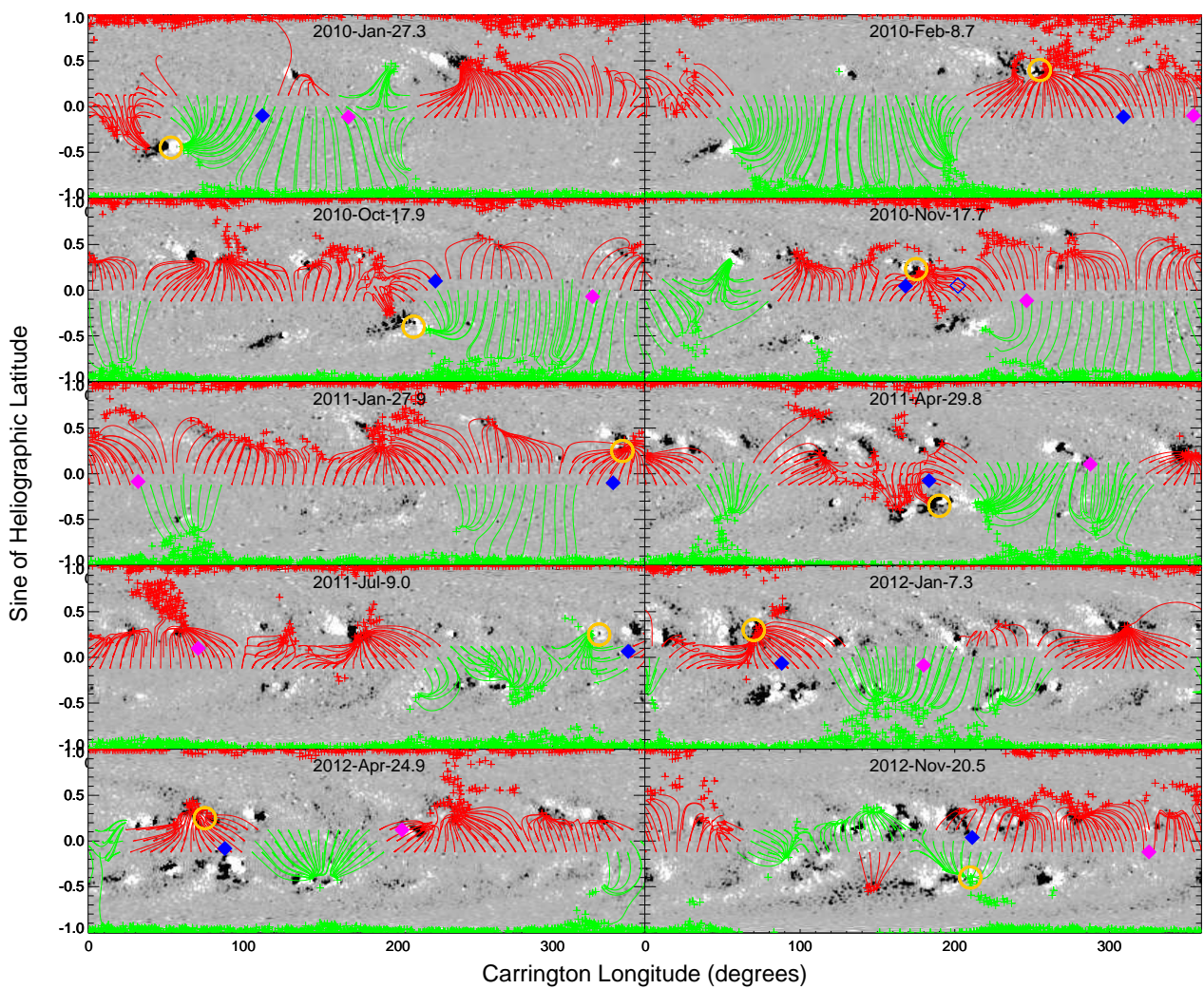}
  \caption{Photospheric magnetic field with PFSS model coronal field lines at the
  start times of the ACE events. Shown are field lines which intersect source surface
  at latitudes 0$^\circ$ and $\pm7^\circ$. Red/green indicates negative/positive
  open field. Blue/purple diamonds mark Earth/STA footpoints on the source surface
  for event start times on ACE. Yellow circles mark the ACE s/c connection locations
  on the Sun. The open blue diamond in panel 2010-Nov-17.7 indicates Earth(ACE) position on 2010-Nov-14.6.
  Note that during the events the s/c footpoint moves several degrees toward the left. }
  \label{wide_fig}
 \end{figure*}

Columns (2)-(3) of Table \ref{table_single} show start times of ACE and STA
$^{3}$He-rich SEP events. The approximate start times were determined from
the increase of number of pulse-height data points above the $^{3}$He background
inspecting the ULEIS/SIT He mass spectrograms. The 385 keV/n $^{3}$He/$^{4}$He event-averaged ratios on ACE and
STA are in columns (4) and (5), respectively. The STA $^{3}$He/$^{4}$He ratios
were corrected for spillover of the $^{4}$He to $^{3}$He \cite{bib:bucik}. Four STA events with $^{3}$He/$^{4}$He
$>$1 were in our earlier study \cite{bib:bucik}. Columns (6)-(7) list the Fe/O
ratios and columns (8)-(9) the $^{3}$He fluences for ACE and STA, respectively.
Column (10) lists ACE and STA s/c common connection regions on the Sun.
The numbers refer to NOAA active regions (AR). According to daily Solar Region Summary, all but two of them
(1197, 1461) were regions with sunspots. Sometimes two $^{3}$He events coming from the same
AR were observed by one s/c. ACE November 14 and 17, 2010 events were
associated with AR 1124. These
two events show quite different (factor of $\sim$14) $^{3}$He/$^{4}$He ratios.
Further examples are STA October 21 and 24, 2010 events associated with AR 1112.

 \begin{figure*}[!htb]
  \centering
   \vspace{-5mm}
  \includegraphics[width=\textwidth]{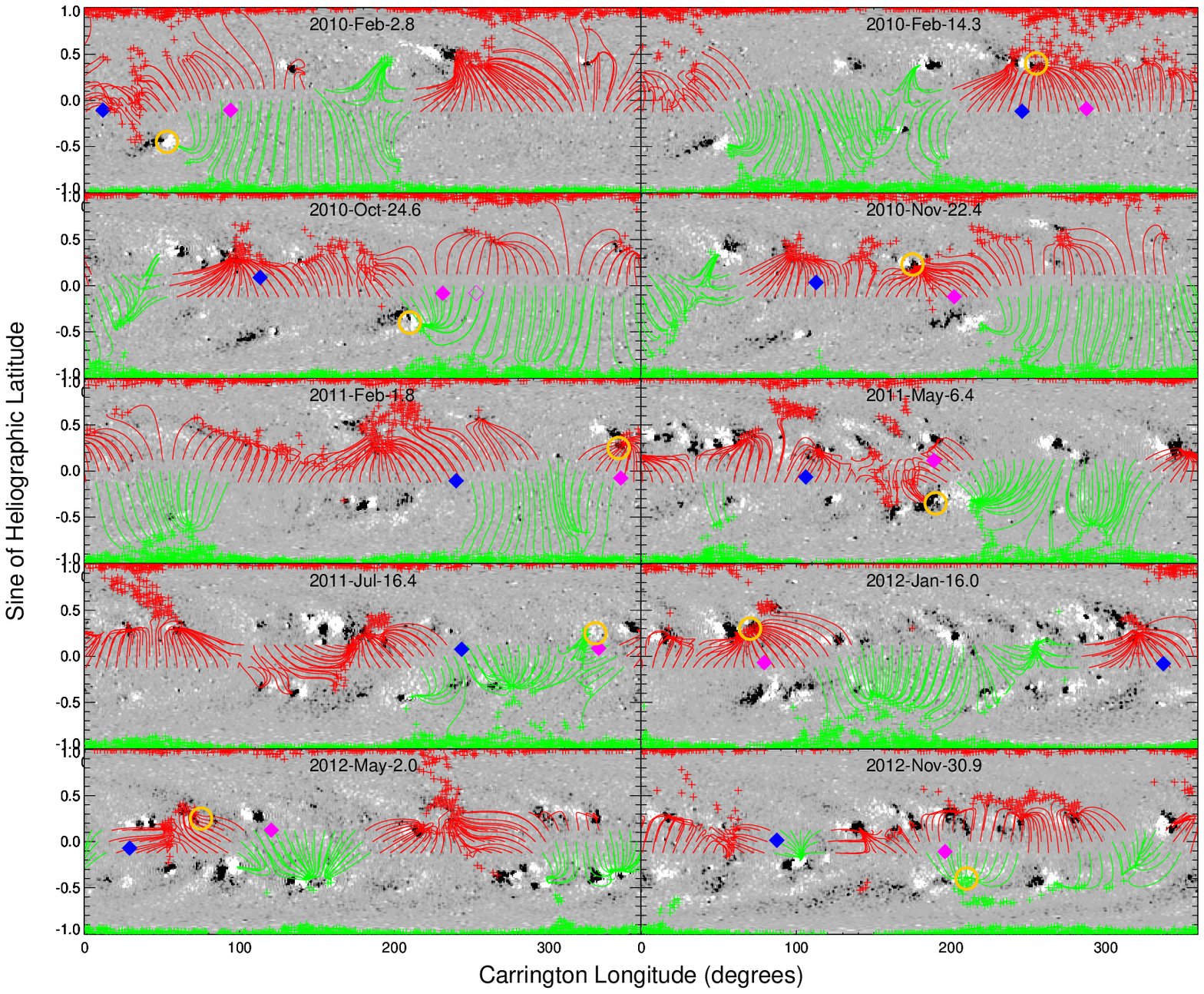}
  \caption{Same as Fig. \ref{wide_fig} but for STA events start times. Open
  diamond in panel 2010-Oct-24.6 indicates STA position on 2010-Oct-21.4. Yellow
  circles mark the STA s/c connection locations on the Sun.}
  \label{wide_fig1}
 \end{figure*}

Figure \ref{wide_fig} shows photospheric magnetic field maps with potential-field source-surface (PFSS)
model field lines at the start times of the ACE events. The field lines shown
are those which are open to the heliosphere for the range of s/c heliolatitudes. This magnetic
field model is based on SOHO/SDO magnetogram data \cite{bib:schrijver} and it is
available via SolarSoft with 6-hour cadence. Blue/purple diamonds mark footpoints of the
Earth/STA at the source surface at 2.5R$_{\odot}$ from Sun center. Outside the
source surface we assume the field is radial and has a form of Parker spiral. The footpoint
longitudes were determined from Parker angle using the measured solar wind speed
with SWEPAM/ACE and PLASTIC/STA instruments.
Yellow circles are source regions on the Sun listed in Table \ref{table_single}. These were identified
by examining the locations of the s/c footpoint during the $^{3}$He-rich periods.
The technique which combines Parker spiral for the interplanetary space and
PFSS model for the corona has been used in identification of sources in the
$^{3}$He-rich SEP events (e.g., \cite{bib:nitta}). In addition to this approach
we also check whether the in-situ magnetic field polarity matches the polarity
from PFSS extrapolations. For this purpose we use 1-hour data from MAG
instruments on ACE and STA.

Figure \ref{wide_fig1} is similar to Fig. \ref{wide_fig} but magnetic field and s/c footpoints
are for STA events start times. We note that PFSS coronal model field lines
are less accurate for STA events. The STA western solar hemisphere was not facing
the Earth during the investigated period and thus not visible by the SOHO or SDO s/c.
Notice on May 2, 2012 event where the PFSS model suggests connection via positive
(green) polarity field while in-situ polarity (not shown) changes sign from
positive to negative at the beginning of the event. The marked AR provides negative
polarity field lines to the distance of $\sim$15$^\circ$ from the STA footpoint.
Offsets of 10$^\circ$ due to limitations of the PFSS model and simple
Parker spiral with constant solar wind speeds have been previously reported \cite{bib:nitta}.

Further we examine if the identified connection regions were flaring before the event.
For ACE we used NOAA Solar Event Reports
(http://www.swpc.noaa.gov) and for STA used SECCHI EUV observations.
We searched for X-ray flares or EUV brightening in 5-8 hour interval before
the event start time. The length of this interval roughly corresponds to
the travel time of 0.2-0.5 MeV/n ions along the spiral with length of 1.1 AU.
Using the NOAA list we found X-ray flares (B or C class) in the connected regions
for 5 out of 11 ACE $^{3}$He-rich SEP events in Table 1 (Feb 8, Oct 17, Nov 14, 2010;
Jan 27, 2011; Jan 7, 2012). For the rest of the ACE events the SECCHI/STA EUV
195~\r{A} images show clear brightening in the connected regions. Note that in the STA
view the ARs for ACE events were seen near the eastern solar limb. The same
solar source for February 8, 2010 ACE event has been reported by another
study \cite{bib:wieden}. During this event, $^{3}$He was simultaneously
observed by the two STEREO s/c but without known source.

 \begin{figure*}[!htb]
  \centering
  \includegraphics[width=0.9\textwidth]{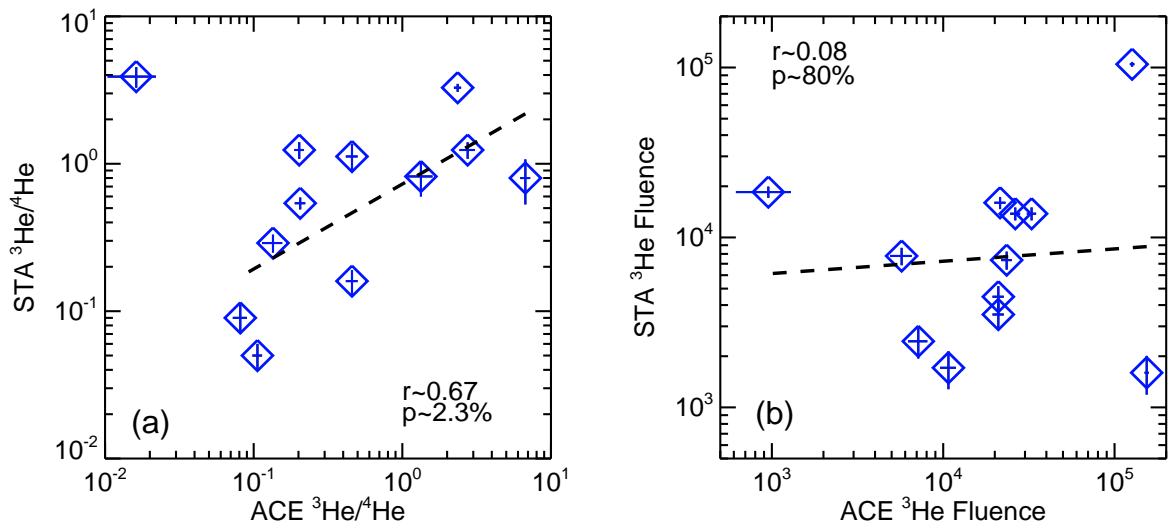}
  \caption{(a) ULEIS/ACE $^{3}$He/$^{4}$He ratio vs. SIT/STEREO-A $^{3}$He/$^{4}$He
   ratio for 320-450 keV/n for events in Table \ref{table_single}. (b) ULEIS/ACE
   $^{3}$He fluence (particles/cm$^2$ sr MeV/n) vs. SIT/STEREO-A $^{3}$He
   fluence for 320-450 keV/n. The dashed lines are power-law fits. The quantities
   $r$ and $p$ denote the correlation coefficient and its statistical significance.}
  \label{simp_fig}
 \end{figure*}

We examined full resolution SECCHI images with 5-minute cadence also for STA events.
The ARs associated with Oct 21, 24; Nov 22, 2010; and Nov 30, 2012 events showed clear
brightening in EUV images. Significant activity with expanding and collapsing
loops was seen in the AR
associated with Feb 1, 2011 event and with material ejections in AR for Feb 14,
2010 event. Only minor brightening was seen in ARs marked for the Feb 2, 2010; May 6,
2011 and Jan 16, 2012 events. Small AR 1246 at the border of the coronal hole, associated
with ACE Jul 09, 2011 event, was very faint on July 16, 2011. SECCHI images showed
two new regions, one emerging on the east side and other on
the west side of the coronal hole. The later one showed only little activity. $^{3}$He
source for May 2, 2012 event is less clear. The marked AR showed no obvious
brightening but the AR on the south with open to ecliptic positive field lines was
highly active with material ejections.

Figure \ref{simp_fig}a shows a scatter plot of 320-450 keV/n $^{3}$He/$^{4}$He ratios on
ACE and STA for events in Table \ref{table_single}. If we exclude the outlier point in the upper left
corner of the plot (April/May 2011 events) the $^{3}$He/$^{4}$He ratio shows significant
positive correlation (0.67) between the two spacecraft with low probability of
being from random population ($<3\%$). This means that the $^{3}$He enrichment did not
change much after $\sim$5-10 days of corotation in different events associated with the same
AR. However, other studies have shown that the $^{3}$He/$^{4}$He ratio changes significantly
from one event to the next event in the same active region \cite{bib:wang}. This figure also shows
large event to event variations in the $^{3}$He/$^{4}$He
ratio spanning about two orders of magnitude.
Figure \ref{simp_fig}b shows scatter plot of 320-450 keV/n $^{3}$He fluences on ACE and STA for
events in Table \ref{table_single}. The $^{3}$He fluences are not correlated even if we exclude the
outlier point in lower right corner (November 2012 events).

\section{Summary}

In the investigated period we have identified several $^{3}$He-rich events successively
observed on ACE and STEREO-A when the two s/c were connected to the same solar region.
Although the sunspot number was lower in 2010 than in 2012, more common sources
were found in 2010, when the s/c were less separated, than in 2012, when separation
between the s/c markedly increased. This could be either because of the decease of
the flaring in the AR or vanishing of the open ecliptic field lines during the
corotation time. Anyway, identifying the events when the s/c are more separated
could set up an upper limit on how long a single source may provide energetic
$^{3}$He ions.

We found that $^{3}$He abundance in the events associated with the same AR does not
show significant temporal changes, although other studies have not shown this.
The largest difference (factor of $\sim$200)
between $^{3}$He enrichments
was seen for ACE Apr 29, 2011 and STA May 6, 2011 events. This is also the case
when the open field lines (of the same polarity) from several ARs were in close proximity
to the s/c footpoint. Other larger differences (factor of 6 and 8) were found for
ACE Nov 14, 2010 and STA Nov 22, 2010 events and for ACE Nov 20, 2012 and STA
Nov 30, 2012 events. We note that at the beginning of the ACE Nov 14, 2010 event
there were open ecliptic lines also from the southern hemisphere AR (not shown
in Fig. \ref{wide_fig}). Thus the cases
with the marked differences are those when multiple sources might contribute
with the $^{3}$He.

The long lasting sources of energetic $^{3}$He identified in this paper tend to be
regions with sunspots. It is not clear how different types/intensities of
the flaring seen in these regions are related to the large event to event
variability in the $^{3}$He enrichment.

\vspace*{0.5cm}
\footnotesize{{\bf Acknowledgment:}{This work was supported by the Bundesministerium
f\"ur Wirtschaft under grant 50 OC 0904. The work at JHU/APL was supported by NASA
under contract SA4889-26309 from the University of California Berkeley. We thank
the ACE SWEPAM/MAG and the STEREO PLASTIC instrument teams for the use of
the solar wind plasma data.}}

\end{document}